\pdfoutput=1

\documentclass[aps,prb,reprint,showpacs,superscriptaddress]{revtex4-1}

\usepackage{graphicx}
\usepackage{graphics}
\usepackage{amsmath}
\usepackage{amssymb}
\usepackage{amsfonts}
\usepackage{dcolumn}
\usepackage{dsfont}
\usepackage{latexsym}
\usepackage{rotating}
\usepackage{color}
\usepackage{latexsym}
\usepackage{bbm}
\usepackage{subfigure}
\usepackage{float}
\usepackage{epsfig}
\usepackage{epsf}
\usepackage{psfrag}
\usepackage{bm}
\usepackage{amsthm}
\usepackage{eucal}
\usepackage{mathrsfs}
\usepackage{url}
\usepackage{braket}
\usepackage{upgreek}
\usepackage{natbib}

\usepackage{color} % use colored text in latex

\usepackage{soul,xcolor}

\usepackage{hyperref}
\hypersetup{
colorlinks=true,final=true,
        linkcolor=blue,
        citecolor=blue,
        filecolor=blue,
        urlcolor=blue,
}
\begin{document}
%\title{Shift photocurrent of transition-metal dichalcogenide monolayers and nanotubes }

%\title{On the origin of the large bulk photovoltaic effect of
%WS$_{2}$ nanotubes}
%\maketitle
\title{Understanding the large shift photocurrent of WS$_{2}$ nanotubes: A comparative analysis with monolayers}
  
\author{Jyoti Krishna}
\affiliation{Centro de F\'{i}sica de Materiales (CSIC-UPV/EHU), 20018, Donostia-San Sebasti\'{a}n, Spain}
\author{Peio Garcia-Goiricelaya}
\affiliation{Centro de F\'{i}sica de Materiales (CSIC-UPV/EHU), 20018, Donostia-San Sebasti\'{a}n, Spain}
\author{Fernando de Juan}
\affiliation{Donostia International Physics Center (DIPC), 20018
Donostia-San Sebasti\'{a}n, Spain}
\affiliation{Ikerbasque Foundation, 48013 Bilbao, Spain}
\author{Julen Ibañez-Azpiroz}
\affiliation{Centro de F\'{i}sica de Materiales (CSIC-UPV/EHU), 20018, Donostia-San Sebasti\'{a}n, Spain}
\affiliation{Donostia International Physics Center (DIPC), 20018
Donostia-San Sebasti\'{a}n, Spain}
\affiliation{Ikerbasque Foundation, 48013 Bilbao, Spain}
  
 \begin{abstract}
We study the similarities and differences in the shift photocurrent contribution to the bulk photovoltaic effect between transition-metal dichalcogenide monolayers and nanotubes.
Our analysis is based on density functional theory in combination with the Wannier interpolation technique for the calculation of the shift photoconductivity tensor.
Our results show that for nanotube radii of practical interest $r>60$~\AA, the shift photoconductivity of a single-wall nanotube is well described by that of the monolayer.
Additionally, we quantify the shift photocurrent generated under realistic experimental conditions like device geometry and absorption capabilities.
We show that a typical nanotube can generate a photocurrent of around 10~nA, while the monolayer only attains a maximum of 1~nA.
This enhancement is mainly due to the larger conducting cross section 
of a nanotube in comparison to a monolayer.
Finally, we discuss our results in the context of recent experimental measurements on WS$_{2}$ monolayer and nanotubes~[Zhang \textit{et al.}, Nature 570, 349 (2019)].
%Our calculations capture the order
%of magnitude and overall shape of 
%the angular current distribution
%observed in a recent experiment on WS$_{2}$~\onlinecite{zhang_enhanced_2019},
%providing a possible explanation for the measured enhancement.

\end{abstract}

\maketitle

\section{Introduction}

The bulk photovoltaic effect (BPVE) offers a promising alternative to traditional solar cells thanks to its ability to generate a d.c.~current upon light absorption in homogeneous materials.
This effect is described as a second-order optical process; hence, it can only take place in 
crystal structures that break inversion symmetry~\cite{sturman_photovoltaic_2021,fridkin_bulk_2001}.
The photovoltage attained in the BPVE
%can surpass the standard Shockley-Queisser limit for the solar-cell
%efficiency, 
%opening the way for devices exceeding current capabilities. Furthermore, the
is not limited by the band gap of the material, giving rise to large measured values~\cite{spanier2016power,osterhoudt2019colossal}.

In the last years, the study of the BPVE, and in particular the shift-current
contribution, has witnessed a 
reinvigorated interest~\cite{young,zhangSwitchable2019,ma2019,yangFlexophotovoltaic2018}. 
While traditionally this effect has been mostly
studied in bulk ferroelectrics such as BaTiO$_3$~\cite{koch1975bulk,young_first_2012}, 
recent theoretical works have emphasized
that the shift current undergoes a significant enhancement 
in two-dimensional (2D) systems such as single-layer
monochalcogenides~\cite{cook2017design,rangel_large_2017}. 
Current efforts include searching for suitable crystal structures
with 2D-like properties, in the hope that they may yield an efficient harvesting of light~\cite{kwak_absorption_2019,bernardi_extraordinary_2013}.

In this context, \emph{nanotubes}, 
which consist of a stack of rolled monolayers, offer an ideal bridge
between a purely 2D system and a bulk crystal structure. 
Early theoretical work by Král and co-workers~\cite{kral_photogalvanic_2000} 
showed the possibility of generating a net shift current in 
acentric and polar BN nanotubes. In addition to the quasiparticle contribution, the role  of excitons 
(collective excitations composed by electron-hole pairs) 
in enhancing  the nonlinear light-absorption process 
has also been addressed~\cite{PhysRevB.103.075402,huang_large_2023}.
Low-dimensional transition metal-dichalcogenides (TMDs) 
also show very good potential as  solar-cell devices due to their 
capacity to absorb a substantial amount of light in the visible range~\cite{bernardi_extraordinary_2013},
and they are also ideal platforms to study van der Walls interactions and
excitonic effects, among other phenomena~\cite{mccreary_-_2018}. 
Interestingly, a recent experiment on
WS$_{2}$ TMD nanotubes reported a 
short-circuit current of around 10~nA~\cite{zhang_enhanced_2019}, yielding
one of the largest figure-of-merit reported to date for nonlinear
processes. 
This remarkable
value may find its origin on the shift-current contribution~\cite{kim2022giant}, 
which is allowed by the
lack of inversion symmetry of these TMD 
polytypes.

In this work we perform a systematic study of the shift current
in WS$_{2}$ monolayer and nanotube structures in order to discern 
the similarities and differences between the two.
Our analysis is based on \textit{ab-initio} density functional theory (DFT) 
in combination with  Wannier-interpolation techniques for an efficient and accurate calculation
of the shift photoconductivity tensor~\cite{ibanez2018ab}.
We find that the optical properties of a single-wall nanotube 
are well described by those of
a monolayer for nanotube radii larger than $\sim60$~\AA, 
which is generally the range of practical interest. 
While the shift photoconductivity of the nanotube is somewhat modified by 
interactions between walls for typical interwall distances,
we do not find a substantial and systematic enhancement.
%such as the one reported in a recent theoretical study~\cite{kim2022giant}.
Despite possessing a similar shift photoconductivity,
we show that a WS$_{2}$ 
nanotube can generate a photocurrent of 
around 10~nA, while the monolayer attains a maximum
photocurrent of order 1~nA.
%The main reason behind this difference is the  
%larger conducting area of a nanotube in comparison to a monolayer.
Finally, we compute the angular current distribution of both nanotube and
monolayer and compare it 
with the one recently measured in experiment~\cite{zhang_enhanced_2019}. 

%In this way, our calculations capture the experimentally observed order
%of magnitude as well as the main shape of 
%the angular current distribution,
%therefore providing a possible explanation for the 
%measured enhancement~\cite{zhang_enhanced_2019}. 

 The paper is organized as follows. In Section~\ref{sec:method} we 
 discuss technical details  regarding the approach to 
 describe the monolayer and nanotube structures and  
 the calculation of 
 optical responses using Wannier interpolation. 
 In Section~\ref{sec:results} we show the bulk of our results;
 after a brief comment on symmetry considerations (Sec.~\ref{sec:symmetry}),
 we analyze the calculated optical photoconductivies of monolayer and
 nanotube structures (Sec.~\ref{sec:photocond})
 and the generated d.c. photocurrent (Sec.~\ref{sec:current}).
 Finally, in Section~\ref{sec:conclusions}, 
 we provide the main conclusions, and supplementary calculations are 
 included in the Appendix. %~\ref{sec:appendix}. 

%%  FIGURE 1
\begin{figure}[]
    \centering
    \includegraphics[width=0.5\textwidth]{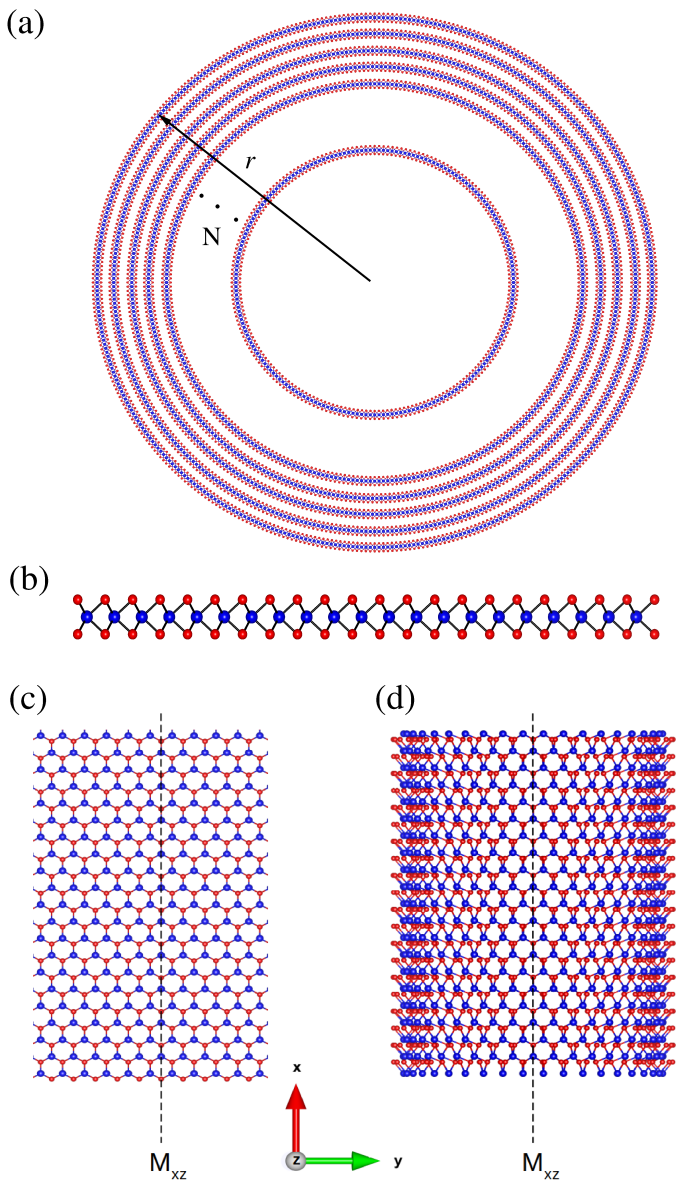}
    \caption{ Side view of two TMD MX$_{2}$ structures: 
    (a) multiwall zigzag nanotube of radius $r$ composed of 
    $N$ layers, and (b) 2D hexagonal monolayer. 
    (c) and (d) show the corresponding top views for monolayer and a single wall 
    nanotube, respectively. The mirror-symmetry plane M$_{xz}$ is denoted by a 
    dashed black colored line. 
    %The coordinate axes marked are in the global coordinate system. 
    The M and X atoms are shown as blue and red balls, respectively.}
\label{fig1}
\end{figure}
%%%%%%%%%%%%%%%%%%%%%%%%%%%%%%%%%%%%

\section{Methods}
\label{sec:method}

We have performed first-principles calculations using DFT 
as implemented in the \textsc{siesta} code package\cite{soler2002siesta}. We have used 
norm-conserving pseudopotentials~\cite{pickett1989pseudopotential,hamann1979norm} 
and we have treated exchange-correlation 
effects by means of the local density approximation~\cite{ceperley1980ground,perdew1981self}.
We have used a basis set centered at the transition metal (M)  and chalcogen (X) 
atoms of the double-zeta type with polarization orbitals, and we have tested 
that the results are virtually unchanged  
using triple-zeta plus polarization orbitals.

We have  considered the trigonal 
 2H-phase crystal structure~\cite{schutte1987crystal} for modelling TMD monolayers of stoichiometry MX$_{2}$. 
 Then, by choosing a three-atom fundamental unit domain (one M sandwiched between two X), 
 we have constructed a single-wall nanotube by rolling up the 2D monolayer along the chiral vector $\vec{C}$ defined as $\vec{C}=n\vec{a}+m\vec{b}$, where $\vec{a}$ and $\vec{b}$ are lattice unit vectors of the monolayer and the chiral (integer) indexes $(n,m)$ determine the chirality of the nanotube. 
 In this work we have focused on the so-called zigzag nanotubes 
 of the type $(n,0)$ [see \ref{fig1}(d)].
For large $n> 15$, the nanotube radius $r$ is proportional to ${n}/{2}$.
In our calculations, we have considered the range  $r\in [10-60]$~\AA, 
and we have incorporated a vacuum region of more than 15~\AA~in every
 non-periodic direction of the computational slab in order to avoid spurious interactions among the periodic images.
 %In our case, the periodicity of the NT is along the tube length (see \ref{fig1}) with lattice constant $a=5.52651 \AA$ for $WS_{2}$. 
 Accordingly, we have sampled the Brillouin zone using a $\Gamma$-centered $\textbf{k}$-mesh of 15 $\times$ 15$\times$ 1 for the monolayer, 10 $\times$ 1$\times$ 1 for the single-wall nanotube, and 5 $\times$ 1$\times$ 1 for the double-wall nanotube with mesh cutoff energy of 100 Ry used in all the calculations.
 %\textcolor{red}{We have checked the convergence of the total 
 %energy with respect to the above parameters with an accuracy of
 %$10^{-5}$}

%%%%%%%  Figure 2
\begin{figure*}
    \centering
    \includegraphics[width=0.5\textwidth]{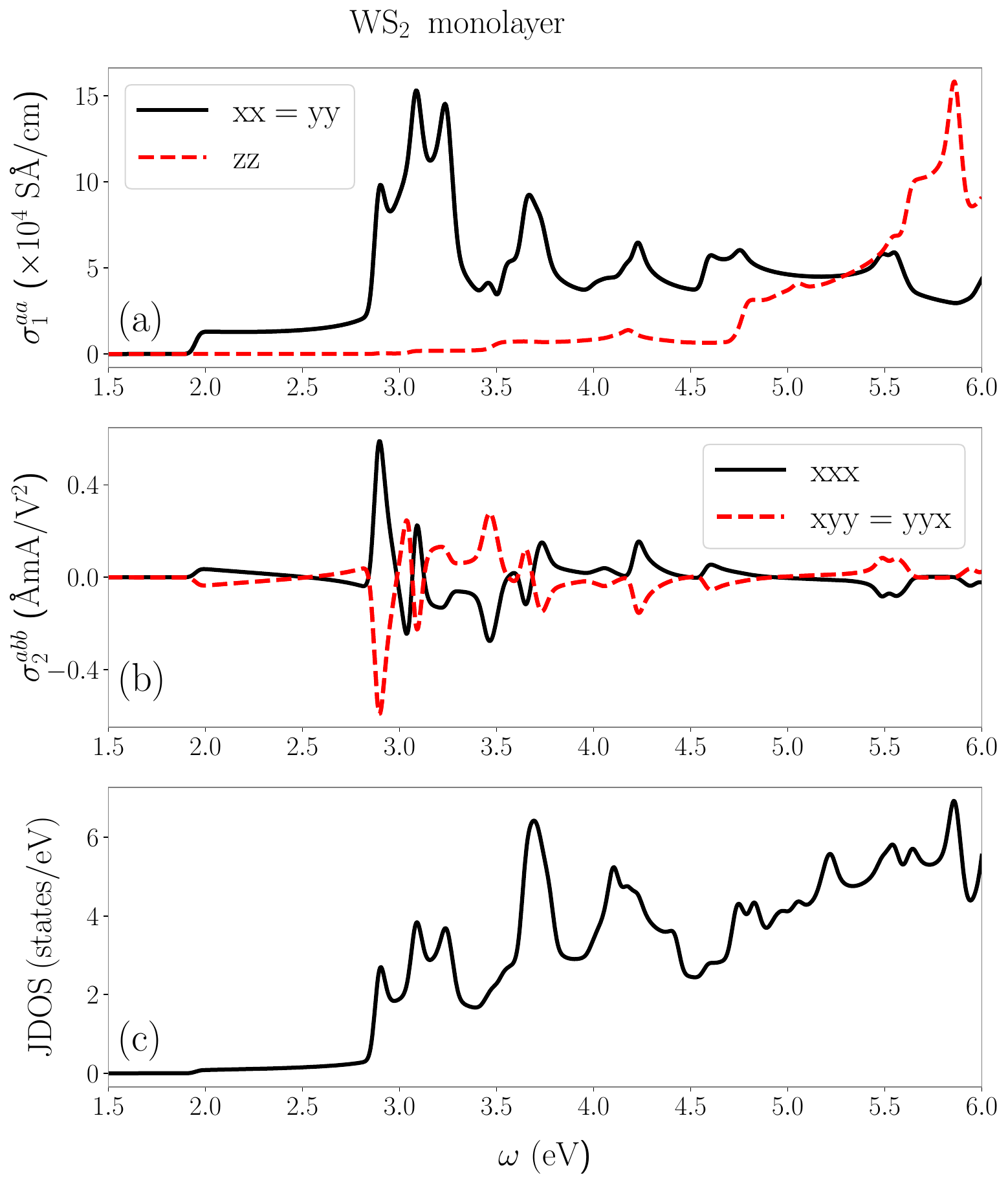}\includegraphics[width=0.5\textwidth]{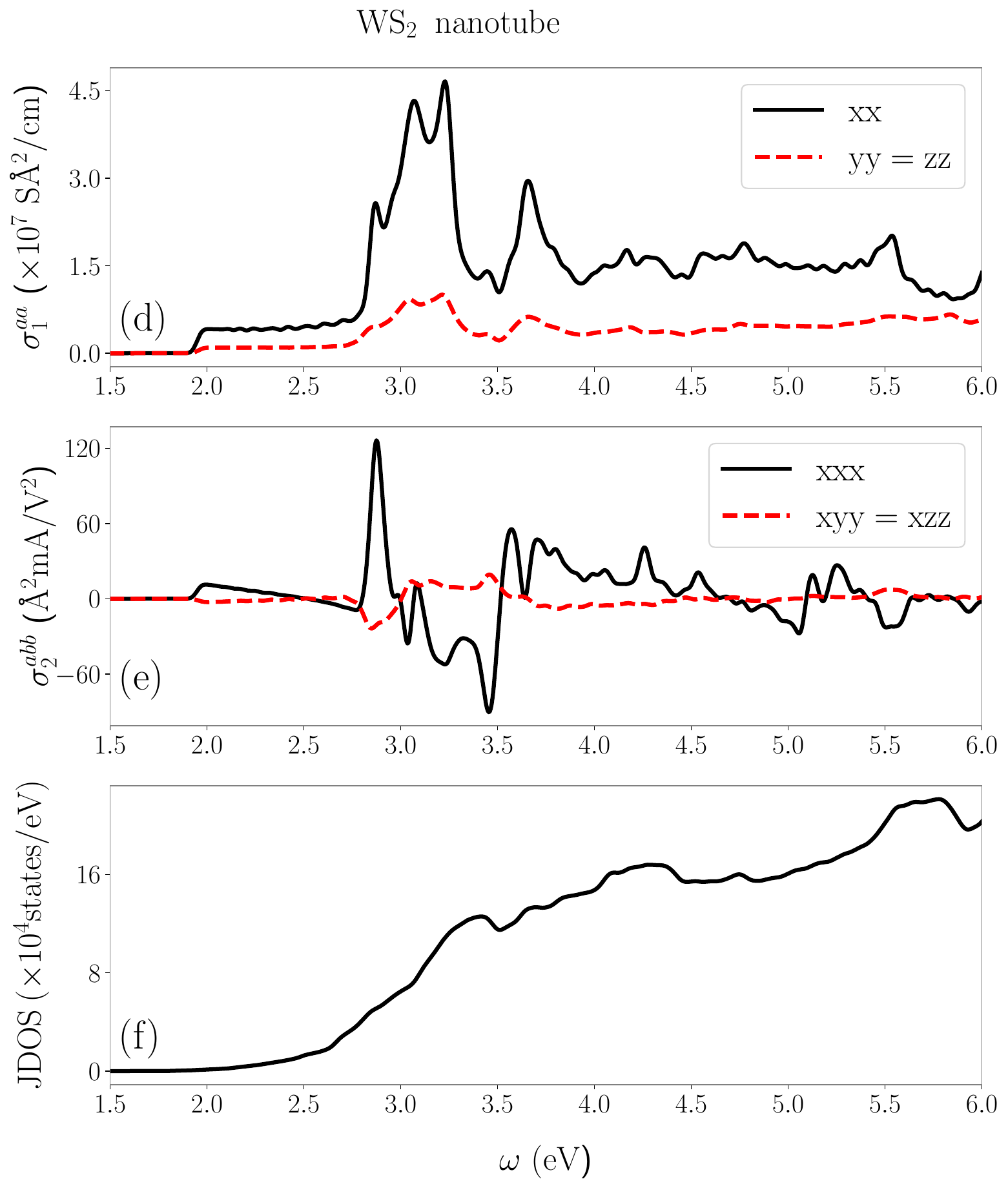}
    \caption{ Symmetry allowed tensor components of linear ($\sigma^{aa}_{1}$) and shift ($\sigma^{abb}_{2}$) photoconductivity versus photon frequency (in eV) of WS$_{2}$ monolayer (panels (a) and (b) respectively) and a single-walled zigzag nanotube of $r=50$~{\AA} (panels (d) and (e) respectively). 
    %The phoconductivities here are renormalized by $R_{m} = 17.1$~{\AA} and $R_{n}^{2} =(125)^{2}$~ {\AA}$^{2}$ for monolayer and nanotube respectively, where, $R_{m}$ and $R_{n}$ represents their corresponding cell-length along non-periodic directions. 
   Panels (c) and (f) show the respective optical joint density of states (JDOS) of monolayer and nanotube. }
    \label{fig2}
\end{figure*}
%%%%%%%%%%%%%%

 In a postprocessing step, we have calculated maximally localized Wannier functions (MLWFs)~\cite{souza2001maximally,marzari1997maximally} from a set of Bloch states, using the \textsc{Wannier90} code package~\cite{pizzi_wannier90_2019}. 
 For the monolayer we have 
 constructed 11 MLWFs comprising 7 high-energy valence bands and 4 low-energy conduction bands using $d$ and $p$ orbitals centered on M and X ions respectively. For the nanotube
 we have constructed the MLWFs by choosing the localized sets of valence and conduction bands around the Fermi level that comprise the $d$ and $p$ orbitals centered on all the M and X ions in the slab, which depends on the chiral index $n$.
 
 Finally, we have computed the linear ($\sigma^{aa}_{1}$)
 and  shift-current ($\sigma^{abb}_{2}$) photoconductivities 
 %\pggt{(for polarized light)}
 using the 
 Wannier-interpolation technique implemented in the \verb|postw90| 
 module~\cite{pizzi_wannier90_2019}. 
 We have computed the dipole matrix element
 and its covariant derivative entering the expression for the
 transition matrix elements~\cite{sipe_second-order_2000} 
 following the approach 
 of Refs.~\onlinecite{wang2006ab}
 and~\onlinecite{ibanez2018ab}, respectively.
 We employed an interpolation $\textbf{k}$-mesh and energy smearing width of 10000 $\times$ 10000 $\times$ 1 and 0.02~eV for the monolayer, respectively, and 1000 $\times$ 1 $\times$ 1 and 0.03~eV for the nanotubes, respectively.

\section{Results}
\label{sec:results}

\subsection{Symmetry considerations}
\label{sec:symmetry}

2D TMD monolayers MX$_{2}$ are formed by a trigonal prismatic network of M transition metal atoms sandwiched by two inequivalent X chalcogen atoms, as illustrated in Fig.\ref{fig1}(b) (side view) and
Fig.\ref{fig1}(c) (top view). The system breaks inversion symmetry but
is symmetric under $y\rightarrow -y$ with a mirror-symmetry plane M$_{xz}$  denoted in Fig.\ref{fig1}(c). 
The point group of the system is D$_{3h}$, 
and the symmetry-allowed components of the shift photoconductivity tensor $\sigma^{abc}$ are $xxx = -xyy = -yyx = -yxy$ (only one independent component).

Regarding the nanotube structure, in this work we will 
report results on the so-called zigzag configuration, as
we have checked that other configurations such as armchair
and chiral ones yield similar or slightly smaller
 optical absorption (this is in line with Ref.~\onlinecite{kim2022giant}).
%given that 
%according to our calculations as well as previous works 
%they show an 
%As illustrated in  Fig.\ref{fig1}(d), in the zigzag nanotube
%the original mirror symmetry of the monolayer is preserved 
%as a family of mirror planes that intersect in the 
%tube axis line.  
%\textcolor{blue}{;these are described by the isogonal point group
%$C_{2nv}$~\cite{damnjanovic2017symmetry}, which for $n>1$ includes the point group  4mm ($C_{4v}$).}
%
A zigzag nanotube belongs to the isogonal point group C$_{2nv}$~\cite{damnjanovic2017symmetry} 
($n$ denotes a positive integer number), which contains rotations around the
tube axis X and vertical mirror planes such as the one illustrated in
Fig.~\ref{fig1}(d). 
%When the radius of the nanotube is sufficiently large, the point group can be approximated as $C_{\infty}$, which for this we can define distinct angular momentum representations ($m=0,\pm 1$) for both the current and electric fields. With a current flowing along the tube-axis, i.e., X, only the representation with $m=0$ remains, 
%This results in symmetry-allowed tensor components for the 
%shift current as $xxx$ and $xyy=xzz$ (two independent components). 
%
The symmetry-allowed tensor components of the shift current are $xxx$ and $xyy=xzz$ (two independent components). 
In practice, this implies that the photocurrent 
flows only along the direction of the tube axis, 
irrespective of the light polarization. 
%Intuitively, this can be understood as a consequence of 
%the broken mirror symmetry along $x-$direction which leads to larger bond asymmetry and hence the photocurrent along that direction.

\subsection{Linear and shift current photoconductivity}
\label{sec:photocond}
\subsubsection{Monolayer and single-wall nanotube}
\label{sssec:mono-swnt}

We begin our analysis by studying the linear and shift
photoconductivity of WS$_{2}$ in various forms. 
Let us start by describing the calculated results for the monolayer, shown
in Fig.~\ref{fig2}(a,b) together with the corresponding 
joint density of states (JDOS) in Fig.~\ref{fig2}(c), 
which provides a measure of 
allowed interband optical transitions~\cite{ibanez2018ab,hu2019dependence}.
As expected, the peaks in the various linear and shift photoconductivity 
spectra coincide with the peaks in the JDOS.
Focusing on the shift current, the maximum value takes place at
$\simeq 2.9$ eV, where it reaches 
$0.57$~{\AA}mA/V$^{2}$. 
As prescribed in Ref.~\onlinecite{cook2017design},
dividing this figure by the
monolayer thickness $l=3.14$ {\AA} 
one can obtain an estimated 
bulk value of about 180~$\upmu$A/V$^{2}$. 
This value is significantly larger 
than the the shift photoconductivity of prototypical ferroelectrics and perovskites\cite{young,zheng2015first,brehm2014first},
and is in line with values reported in other 2D monolayers 
such as GeS~\cite{rangel_large_2017}.

We turn now to analyze the optical properties of a single-wall
nanotube of radius $r=50$ {\AA}. 
The calculated band gap $\simeq~$1.9 eV is very close to 
that of the monolayer value
(see electronic band structures in  Appendix),
in line with previously reported values~\cite{staiger2012excitonic,zibouche2012layers}. 
For both the linear [Fig.~\ref{fig2}(d)] and shift current [Fig.~\ref{fig2}(e)] spectra, the dominant 
photoconductivity component 
corresponds to the tube axis $\sigma_{1}^{xx}$ and $\sigma_{2}^{xxx}$,
respectively. 
In both cases, the shapes are very similar to the associated 
components calculated for the monolayer, and the maximum
shift current reaches
120~{\AA}$^{2}$mA/V$^{2}$. 
Note, however, that due to the difference in units 
between 
Fig.~\ref{fig2}(b) and Fig.~\ref{fig2}(e), a 
direct quantitative comparison
between monolayer and nanotube shift photoconductivity is not straightforward. 

%%%%%%  FIGURE 3  %%%%%%%%%%%%%%%%%%%%%%%%%%%
\begin{figure}[t]
    \centering
    \includegraphics[width=0.5\textwidth]{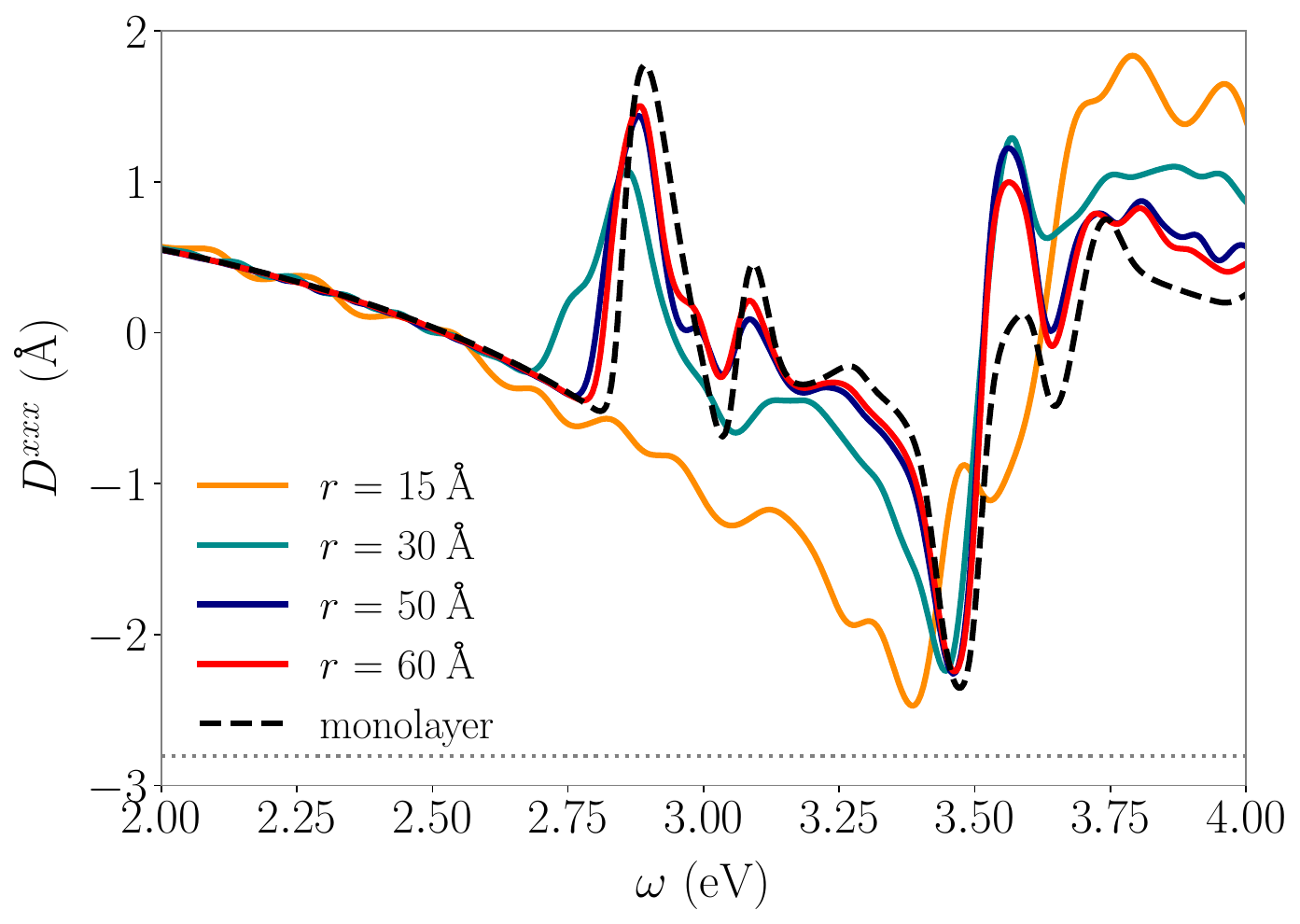}
    \caption{A plot comparing the shift distance tensor ($D^{xxx}$) as a function of photon frequency ($\omega$ in eV) for nanotubes with varying radii ($r$) and a monolayer. The monolayer's result is represented by a black dashed line, while the nanotubes are indicated by colored lines. The grey dashed line shows the average bond length of $2.8$~\AA~between the ions in the monolayer.}
    \label{fig3}
\end{figure}
%%%%%%%%%%%%%%%%%%%%%%%%%%%%%%%%%%%%%%%%%%%%%%

In order to overcome this subtlety, we have opted for considering 
a quantity known as the
\emph{shift distance}. This magnitude
quantifies the real-space distance travelled by electronic carriers 
upon photoexcitation as a consequence of the shift-current mechanism\cite{nastos2006optical}.
It is defined as 
\begin{equation}
\label{eq:shift-distance}
\centering
    D^{abb} = \frac{\hbar}{\epsilon_{0} e} \times \frac{\sigma_{2}^{abb}}{\text{Im} \epsilon^{bb}},
\end{equation}
 where $\epsilon^{bb}= 1+ i\sigma^{bb}_{1}/\epsilon_{0}\omega$ is the complex dielectric function within the independent-particle approximation, with $\epsilon_{0}$ the vacuum permittivity and $e$ as electron charge.
 Since Eq.~\ref{eq:shift-distance} involves a ratio between the
 quadratic and linear absorption coefficients, the shift distance has
 always length units, 
 allowing a direct comparison of monolayer and nanotube results.
 In Fig.~\ref{fig3} we show the calculated results of the main
 component $D^{xxx}$ for the monolayer as well as for single-wall
 nanotubes for radii ranging from 15 to 60 \AA. 
 While the peak structure closely follows that of optical properties in all cases, 
 the shape is significantly altered, showing maxima at
 $\sim 3.5$ eV. Overall, the calculated shift distance 
 is of the order of  
 the average bond length between S$-$S and W$-$S atoms in the
 monolayer configuration, indicated in the figure.
 This magnitude is  in line with what has been 
 previously reported in bulk materials~\cite{nastos2006optical,ibanez2020directional}.

 We  end this section by inspecting the dependence of 
the shift distance on the nanotube radius. 
 For a nanotube with $r=15$~\AA, we find no peak at $\sim2.9~\mathrm{eV}$
 and the shape of 
 $D^{xxx}$ deviates  significantly from 
 the monolayer result. 
 However, as $r$ is increased the nanotube shift distance 
 tends to match the shape as well as magnitude of the monolayer
 result; for 
 $r=60$~\AA, only slight deviations are visible.
 This is an important result, as it shows that 
 nanotubes with radius larger than
 60~\AA~fall within the monolayer limit. Given that 
 in reality, all walls of synthesized nanotubes have $r>$60~\AA\cite{zak2010scaling,qin2017superconductivity},
 their optical properties can be conveniently described by 
 those of the monolayer, provided interwall interactions
 are not too large. The latter are analyzed in the following section.

\subsubsection{Double-wall nanotube}
\label{sssec:doublewall}

%%%%%%%%%%%%%%%FIGURE 4 %%%%%%%%%%%%%%%%%%%%%%

\begin{figure*}[t]
    \centering
    \includegraphics[width=1.0\linewidth]{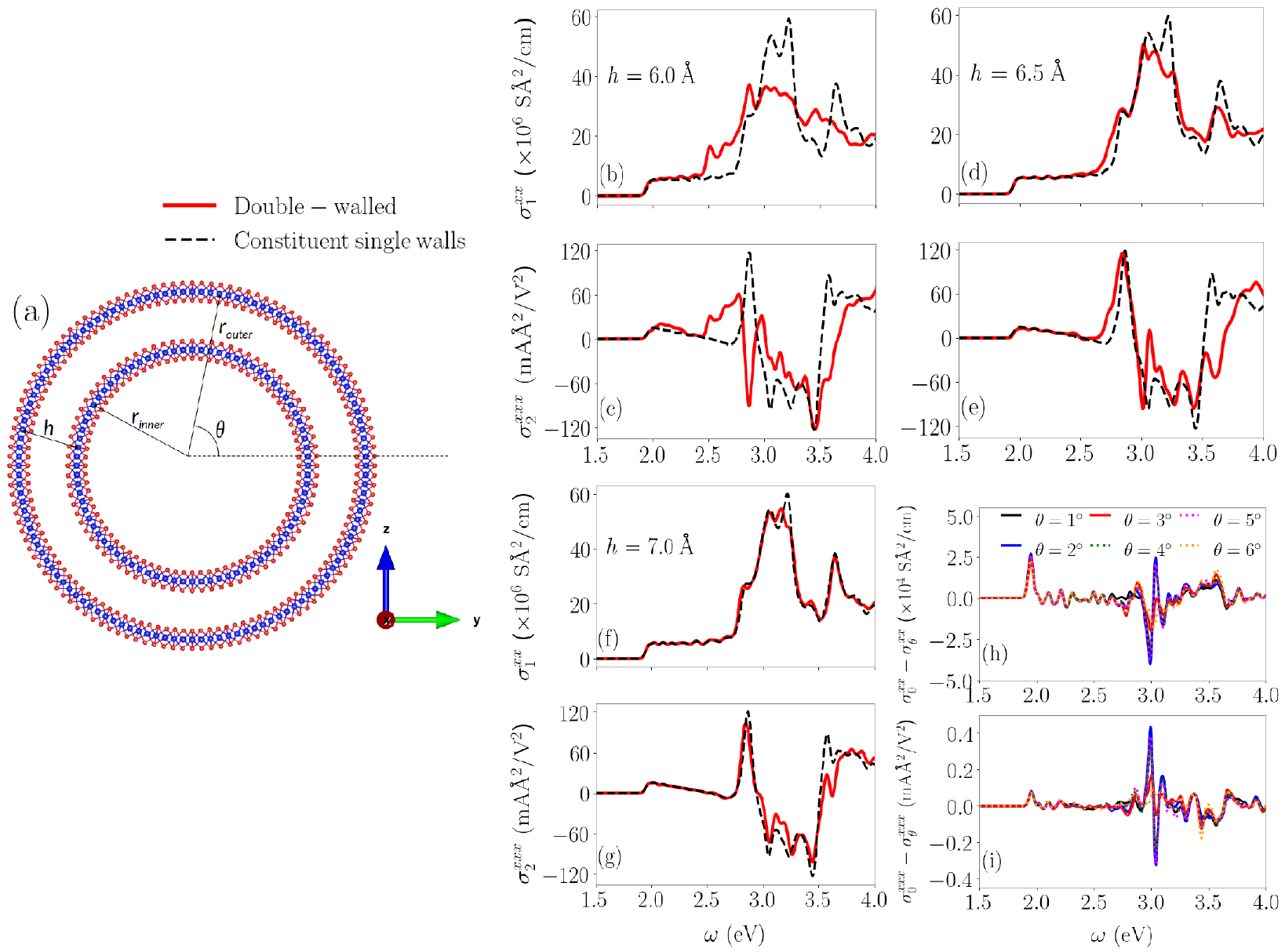}
    \caption{(a) Side view of a double-wall nanotube defined by its inner radius ($r_\text{inner}$), outer radius ($r_\text{outer}$), and the distance between the walls ($h$). The chirality of the inner wall is fixed at (60,0) and we varied the outer in the range (72,0), (73,0) and (74,0) to maintain interlayer distances of 6.0~\AA, 6.5~\AA, and 7.0~\AA, respectively. The angle $\theta$ signifies the rotational displacement of the outer wall relative to the inner wall. Figures (b)-(g) show the linear and shift current photoresponses of the double-wall nanotubes having interwall distances of $h\in [6,6.5,7]$~\AA~(red solid lines). These are compared with responses of the individual single-wall nanotubes constituting a double wall (black dashed lines). Figures (d) and (e) show the photoresponse for  $h = 6.5~\text{\AA}$ which is closest to the interwall distance of WS$_{2}$ nanotubes~\cite{zak2010scaling,bruser2014single}. Figures (h) and (i) depict how the photoresponses are affected when the outer wall is rotated with respect to the inner wall by angle $\theta$ in range $\in[1^{\circ}-6^{\circ}]$ (for higher angles, the structure repeats). Here, we have shown the results for a smaller double wall nanotube with $r_\text{inner}$ = 30~\AA, $r_\text{outer}$ = 37~\AA, and $h$ = 6.5~\AA.  }
    \label{fig4}
\end{figure*}
%%%%%%%%%%%%%%%%%%%%%%%%%%%%%%%%%%%%%%%%%%%%%%
%%%%%%%%%%%%%%%%%%%%%%%%%%%%%%%%%%%%%%%%%%%%%%

Here we report on the linear and shift 
photoconductivity of a double-wall 
WS$_{2}$ nanotube; see Fig.~\ref{fig4}(a) for a schematic illustration. For their construction, we stacked two single-wall zigzag nanotubes of different radius on one another.
In our calculations, we have fixed the radius of the inner wall 
at $r=30$~\AA~ and varied the radius of the outer wall 
in order to sample the photoresponse for varying
interwall distance $h$.
The interwall 
distance in the ideal WS$_{2}$ bilayer
structure is $h\approx6.23$~\AA,
which increases by $\sim 2\%$ in nanotubes~\cite{zak2010scaling,bruser2014single}.
Having this in mind, 
in Figs.~\ref{fig4}(b)-(e) we show the calculated $\sigma^{xx}_{1}(\omega)$ 
and $\sigma^{xxx}_{2}(\omega)$ for $h=$6~\AA, 6.5~\AA~and 7~\AA.
In addition to the double-wall results, we have also included 
results corresponding to the 
individual constituent walls, hence
the difference between the two sets can be attributed to the 
interwall interaction.\\

Figs.~\ref{fig4}(b)-(c) for $h = 6$~\AA~show that, while 
the interwall interaction induces visible deviations from the single wall result, it does not
alter the order of magnitude of the optical responses.
The largest effect on the shift current takes place at 2.9 eV, where  
peak of the double-wall result
flips the sign. 
However,  already for $h= 6.5$~\AA~[Figs.~\ref{fig4}(d)-(e)] the interwall interaction 
induces  only minor differences with respect to 
the single wall result, and the main spectral features are practically
restored. 
Given that synthesized WS$_2$ nanotubes are likely to have an 
interwall distance closer to 6.5~\AA~than 
6.0~\AA,
our calculations indicate that the interwall interaction
does not   affect the photoresponse properties
to the extent that it could explain the 
enhancement reported in the experiment 
of Ref.~\onlinecite{zhang_enhanced_2019}. 
We have verified that this holds for different
rotation angles of one wall with respect to the other
[see Fig.~\ref{fig4}(h)-(i)] as well as for other values of the inner and outer walls (keeping the considered
range of interwall distance).

The above results are in apparent contrast to some of the results reported in Ref.~\onlinecite{kim2022giant},
where an acute enhancement of the shift photoconductivity
(but not of the linear absorption) was observed in double-wall nanotubes with interwall distance around $6$~\AA, which was attributed to a wall-to-wall charge shift. 
We have not found evidence of this enhancement in our calculations, even when using the same radii reported in Ref.~\onlinecite{kim2022giant}. 
We note that, unlike in the theoretical approach employed in
Ref.~\onlinecite{kim2022giant}, we do not resort to a tight-binding model
derived from Wannier functions as we keep the whole matrix structure of 
both the Hamiltonian as well as the position matrix elements~\cite{ibanez2018ab}. 
This might explain part of the difference with the results of that work,
given that position matrix elements can play an important role in
the shift-current generation~\cite{ibanez2022assessing}.

%%%%%%  FIGURE 5  %%%%%%%%%%%%%%%%%%%%%%%%%%%
\begin{figure}[]
    \centering
    \includegraphics[width=0.5\textwidth]{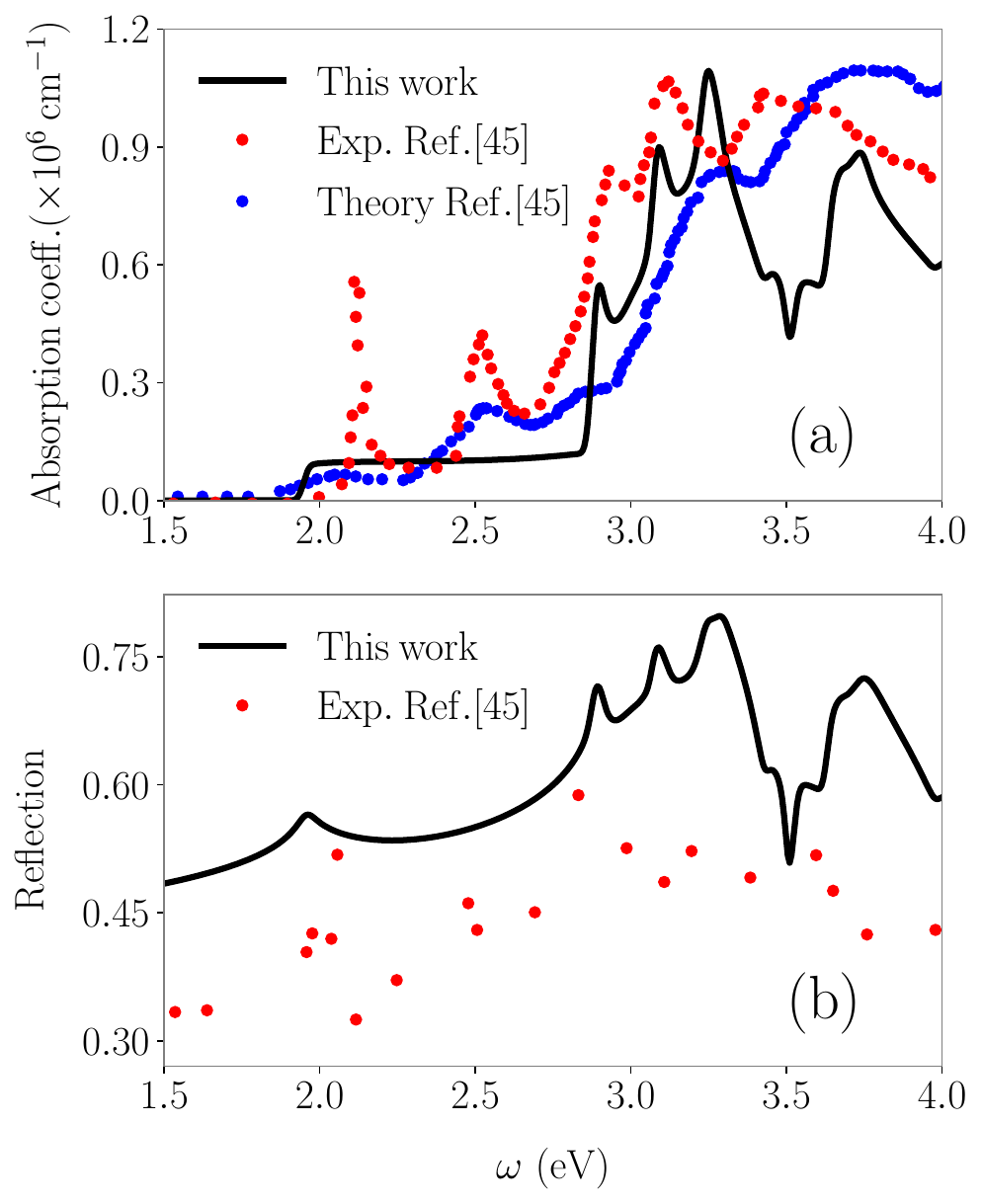}
    \caption{The plot displays the (a) absorption coefficient and (b) reflection ratio as a function of photon frequency $\omega$ (in eV). The solid black line represents our theoretical calculations for the WS$_{2}$ monolayer. For comparison with our results, the experimental and theoretical data from Ref.~\onlinecite{liu_temperature-dependent_2020} are displayed in the figures as indicated in red and blue dots respectively. 
    The sharp peaks at $\sim$2.2 eV and
    $\sim$2.5 eV correspond to excitons measured in experiment.}
    
    \label{figRA}
\end{figure}
%%%%%%%%%%%%%%%%%%%%%%%%%%%%%%%%%%%%%%%%%%%%%%

%%  FIGURE 6 
\begin{figure}
    \centering
    \includegraphics[width=0.5\textwidth]{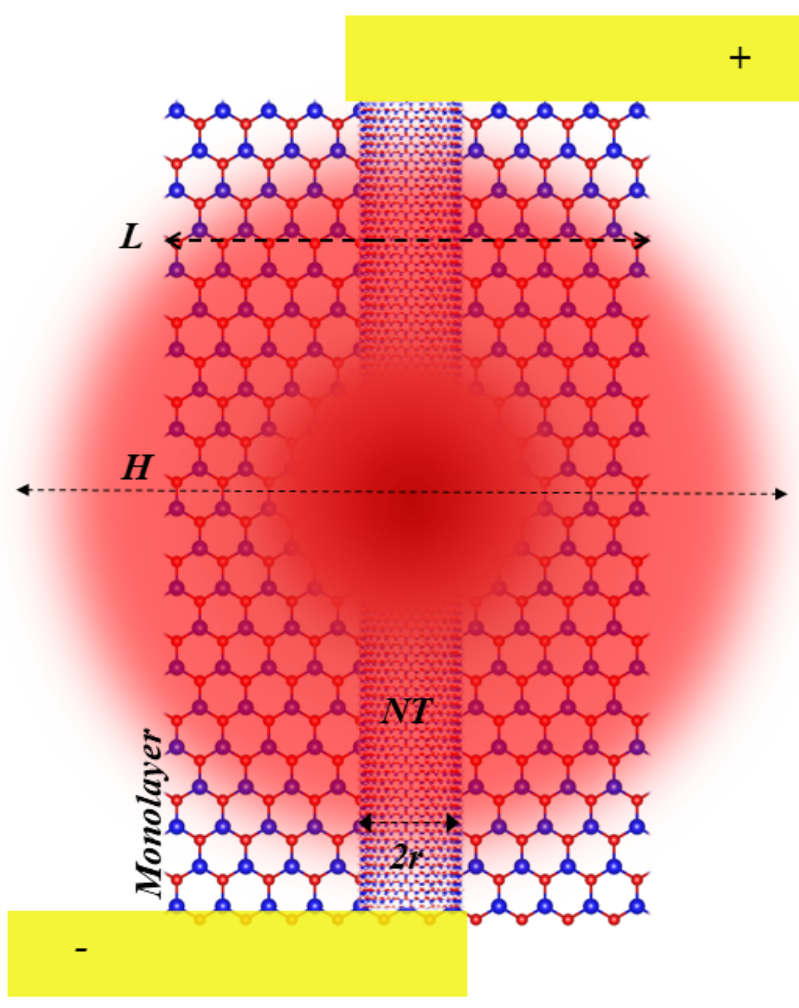}
    \caption{Schematic diagram showing the experimental setup which we used for estimating the total current flowing through monolayer and nanotube. We have denoted the spot size of the laser profile as $H$, which is illuminated on a monolayer of width $L$, and a nanotube of radius $r$. The current flows between the electrodes, which are marked by the terminals +/-.  }
    \label{fig6}
\end{figure}
%%%%%%%%%%%%%%%%%

\subsection{Total current: monolayer vs nanotube}
\label{sec:current}
\subsubsection{Estimates of relevant quantities}

%Experimental absorption coefficient MoS2, $\alpha^{-1}\sim 100$ \AA~\cite{kwak_absorption_2019,bernardi_extraordinary_2013}

We turn now to analyze the factors involved 
in the generation of the total shift current of WS$_{2}$ monolayer and nanotube.
%
%Our estimates are based on the calculated photoconductivities described
%in Sec.~\ref{sssec:mono-swnt}. 
%Motivated by the double-wall nanotube 
%results of Sec.~\ref{sssec:doublewall}, we assume
%that the interwall interaction does not substantially modify the
%shift photoconductivity of the composing layers.
%Furthermore, given that all layers in a real nanotube
%have radius $r>60$\AA, they fall within the monolayer limit and
%therefore we approximate the photoconductivity of every layer by that
%of the monolayer.
%
For a material with thickness $d$,
the  shift current generated under linearly polarized light $\mathbf{E}$
in a direction normal to the incidence can be written as~\cite{PhysRevB.101.045104}
 \begin{equation}
 \centering
 {J}_{a} =G^{abb}(\omega)\cdot[1-R(\omega)]
 \cdot(1-e^{-\alpha_{bb}(\omega)\cdot d})
 \cdot w\cdot I_{b},
 \label{eq:totalI}
  \end{equation}
   with $I_{b}=c\epsilon_{0} E^{2}_{b}/2$ and $c$ the speed of light.
  There are several quantities entering the above expression, which we 
  now discuss one by one in the context of the experimental setup of Ref.~\onlinecite{zhang_enhanced_2019}.

The factor $1-R(\omega)$ describes the portion of light that is not
reflected at the surface between the vacuum and the material. It involves
the \emph{reflectivity}
\begin{equation}\label{eq:relfectivity}
    R(\omega)=\dfrac{[1-n(\omega)]^{2}+\kappa(\omega)^{2}}{[1+n(\omega)]^{2}+\kappa(\omega)^{2}},
\end{equation}
where the coefficients $n$ and $\kappa$ are related to the 
real (R) and imaginary (I) parts of the 
complex dielectric function $\epsilon^{bb}=\epsilon^{bb}_{\text{R}}+i\epsilon^{bb}_{\text{I}}$ 
as $\epsilon^{bb}_{\text{R}}=n^{2}-\kappa^{2}$ and 
$\epsilon^{bb}_{\text{I}}= 2n\kappa$.
%$\epsilon^{bb}=\epsilon^{bb}_{\text{R}}+i\epsilon^{bb}_{\text{I}}\equiv 1+4\pi i\sigma^{bb}_{1}/\omega$ 
In Fig.~\ref{figRA}(b) we show the calculated reflectivity factor $R(\omega)$.
It shows that at the band-edge approximately half the incoming light is reflected,
whereas, at the peak energy $\omega\simeq 2.9$ eV approximately 70$\%$ of light is 
reflected; both these values are in rather good agreement with 
experimental measurements of Ref.~\onlinecite{liu_temperature-dependent_2020},  
which we have reproduced in the figure.
Given that the reflectivity is mainly a surface property, we assume that 
the same factor applies to the monolayer and the nanotube. 
This is backed up by a recent 
experiment on WS$_{2}$ nanotubes and 2D sheets, which displays very similar 
levels of reflection intensity for most frequencies\cite{liu2019ws2}.

The factor $I_{b}$ in Eq.~\ref{eq:totalI} accounts for the intensity of the electric
field.  Ref.~\onlinecite{zhang_enhanced_2019} employed a Gaussian beam, and  
for a laser of 632.8 nm wavelength, the average electric field 
strength (or power density) over the spotsize is $1.39\cdot 10^{4}$ W/cm$^{2}$.
This number describes appropriately the action of the electric field
over the monolayer, given that its dimensions 
described by $L=2$~$\upmu$m are of the order of the 
spotsize diameter $H=2.5$~$\upmu$m~\cite{zhang_enhanced_2019}
(see Fig.~\ref{fig6} for a sketch).
In turn, the effective strength of the electric field acting on
the nanotube is larger than the average value, given that it is placed in the middle
of the laser spotsize where light intensity is highest 
and its radius $r=90$ nm is small in comparison to $H$ 
(see Fig.~\ref{fig6}). Considering 
the Gaussian profile employed in Ref.~\onlinecite{zhang_enhanced_2019},
we find that the effective strength in the nanotube is approximately $15\%$
larger than the average value, hence, we use 
$I_{b}=1.6\cdot 10^{4}$ W/cm$^{2}$ for the nanotube. 

%{\color{red}FJ: $W/cm^2$ are units of intensity, not electric field squared, which should be $(V/cm)^2$ There is the usual factor with epsilon and c here?}

Finally, we come to analyze the geometric and light-absorption terms
in Eq.~\ref{eq:totalI}. 
  The symbol $w$ denotes the length of the material exposed to light illumination,
which is $w_{\text{M}}=L$ in the case of the monolayer
and $w_{\text{NT}}=2r$ for the nanotube (see Fig.~\ref{fig6}).
Plugging the numbers we get that 
$w_{\text{M}}\simeq 10\cdot w_{\text{NT}}$, reflecting the fact that
the monolayer is much wider than the nanotube diameter.
  On the other hand, 
$G^{abb}(\omega)$ in Eq.~\ref{eq:totalI} stands for the \emph{Glass coefficient}~\cite{glass1974high,tan2016shift}
\begin{equation}\label{eq:Glass}
    G^{abb}(\omega)=\dfrac{2\sigma_{2}^{abb}(\omega)}{c\epsilon_{0}\sqrt{\epsilon_{r}}\cdot\alpha_{bb}(\omega)},
\end{equation}
with $\epsilon_{r}$ the dielectric constant of the material.
The Glass coefficient thus involves the ratio between the shift current and the \emph{absorption coefficient}~\cite{PhysRevB.101.045104}
 \begin{equation}\label{eq:alpha}
\alpha_{bb}(\omega) =\sqrt{2}\dfrac{\omega}{c}
\sqrt{\left|\epsilon^{bb}\right|-\epsilon^{bb}_{\text{R}}}
    \end{equation}
%with $\epsilon^{bb}=\epsilon^{bb}_{\text{R}}+i\epsilon^{bb}_{\text{I}}\equiv 1+4\pi i\sigma^{bb}_{1}/\omega$ 
%the complex dielectric function.
The inverse of the absorption coefficient  
describes the light penetration depth into the material. We note that, in
the limit of thin materials, 
$\alpha_{bb}^{-1}(\omega)>>d$, and therefore, the expression of Eq.~\ref{eq:totalI}
reduces to ${J}_{a} =\sigma^{abb}_{2}(\omega)\cdot[1-R(\omega)]\cdot d
 \cdot w\cdot I_{b}/\sqrt{\epsilon_r}$, which is independent of $\alpha_{bb}(\omega)$
and involves the cross section $d  \cdot w$ normal to the flow of current.

In Fig.~\ref{figRA}(a) we show the calculated absorption coefficient for WS$_{2}$.
The figure shows that $\alpha_{bb}(\omega)$ ranges between 
$\sim$1$\cdot10^{5}$~cm$^{-1}$ at the band-edge and 
$\sim$5$\cdot10^{5}$~cm$^{-1}$ at the peak
energy $\omega=2.9$ eV. These values are in good agreement with previously reported
experimental measurements 
and theoretical estimates of $\alpha_{bb}(\omega)$~\cite{liu_temperature-dependent_2020}.
In practical terms, this means that the light penetration depth ranges between
$\simeq$1000 $\mathrm{\AA}$ at the band-edge and 
$\simeq$200 $\mathrm{\AA}$ at $\omega=2.9$ eV. 
In the case of the monolayer, its thickness $d_{\text{M}}=3.14$~$\mathrm{\AA}$
is orders of magnitude smaller than the penetration depth.
As for the nanotube, it is typically composed of $\sim$25 layers
and light traverses them twice in most regions. 
%with interwall distances of the order of 6 $\mathrm{\AA}$ %peio
Considering the interwall 
distance of $\sim$6.5 $\mathrm{\AA}$~\cite{zak2010scaling,bruser2014single}, 
a nanotube is roughly $d_{\text{NT}}\simeq 300$ $\mathrm{\AA}$ thick;
given that $d_{\text{NT}}\lesssim\alpha^{-1}_{bb}(\omega)$,
most layers of the nanotube are active
in absorbing light. 
This in turn means that $d_{\text{NT}}\simeq 100\cdot d_{\text{M}}$,
reflecting the fact that a nanotube is much ``thicker'' than a monolayer.
Combining with the width factor discussed earlier,
we conclude that 
$w_{\text{NT}}\cdot d_{\text{NT}} \simeq 10 \cdot w_{\text{M}}\cdot d_{\text{M}}$,
which represents roughly an order of magnitude enhancement
of the nanotube as compared to the monolayer.

%The above is an important result: considering the setup 
%of Ref.~\onlinecite{zhang_enhanced_2019},
%Eq.~\ref{eq:Icomp} indicates that a zigzag nanotube 
%generates at least a hundred times more shift current
%than a monolayer
%under the assumption of equal photoconductivity in the two structures.

\subsubsection{Angular dependence and magnitude of the photocurrent}

%%%%%%%%  fig 7 %%%%%%%%%%%%
\begin{figure*}[t]
    \centering
    \includegraphics[width=0.5\textwidth]{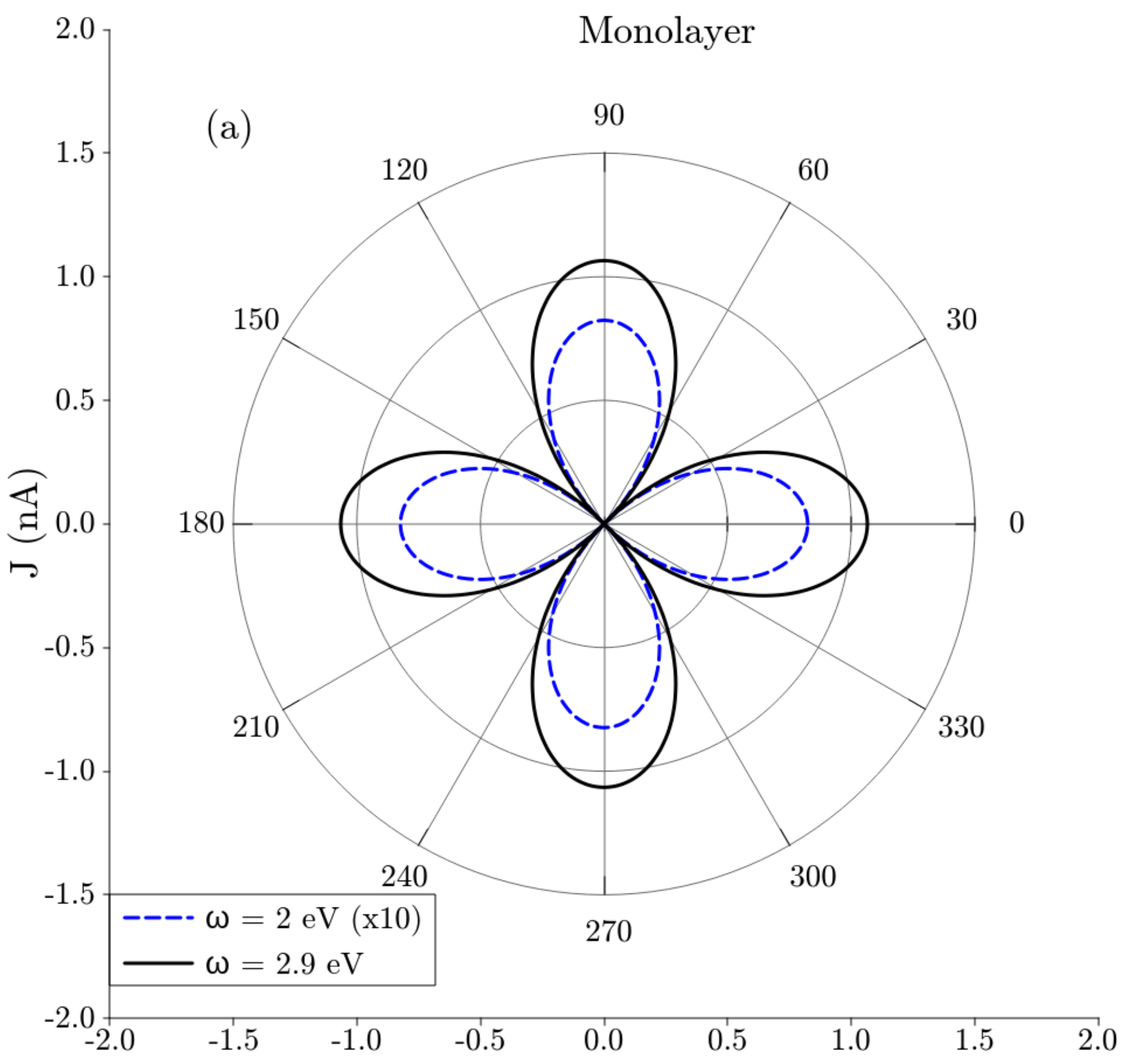}\includegraphics[width=0.5\textwidth]{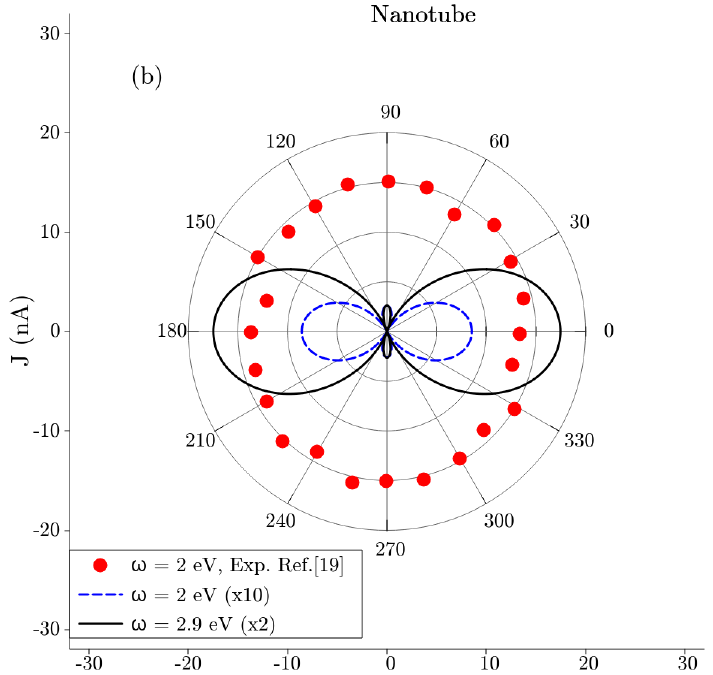}
    \caption{Polar plot showing the variation of shift current (J) in nA as a function of the polarization angle $\theta$ at normal incidence for WS$_{2}$ (a) monolayer and (b) nanotube. The excitation energies $\omega$=2 eV and 2.9 eV correspond to the response at the band-edge and peak respectively. The nanotube's result is compared with the experimental data of device1 ($r = 90$ nm)\cite{zhang_enhanced_2019} (red dots).}
   \label{fig7}
\end{figure*}
%%%%%%%%%%%%%%%%%%%%%%%

As the last step in our analysis, we 
study the angular dependence of the 
shift photocurrent in the two structures based on
the arguments of the preceding sections, and quantitatively 
compare our results to the experimental measurements of 
Ref.~\onlinecite{zhang_enhanced_2019}.
We assume linearly polarized light in the $xy$ plane 
under normal incidence, with  
its electric field described by
\begin{equation}
 \mathbf{E}(\omega) = E(\omega) (\cos \theta \hat{i} + \sin \theta \hat{j}).
\end{equation}

The total current generated along the $x$ axis
can be then expressed as
\begin{equation}
    J^{x}(\theta,\omega) = A\cdot (G^{xxx}(\omega) \cos^{2}\theta + G^{xyy}(\omega) \sin^{2}\theta), 
\end{equation}
where $A=[1-R(\omega)]
 \cdot(1-e^{-\alpha_{bb}(\omega)\cdot d})
 \cdot w\cdot I_{b}$  gathers the remaining factors of  Eq.~\ref{eq:totalI}.

%\begin{equation}
%    I^{y}_{m}(\theta,\omega) = (\sigma^{yyy}(\omega) sin^{2}\theta + \sigma^{yxx}(\omega)  cos^{2}\theta ) C_{2}
%\end{equation}

To illustrate our results, 
in Fig.~\ref{fig7}  we have considered two photon energies; 
2~eV, where Ref.~\onlinecite{zhang_enhanced_2019} measured 
maximum current, and 2.9~eV, where our 
calculations predict maximum shift photoconductivity (see Fig.~\ref{fig2}).
%The laser excitation wavelength used in the experiment ($632.8 nm$) falls approximately near our calculated band edge of NT and the monolayer. 
Let us begin by discussing the monolayer results
of Fig.~\ref{fig7}(a). 
Our calculations predict a photocurrent smaller than 0.1~nA 
for photon energy 2~eV, while at 2.9~eV it is maximum 
and of the order of $1$~nA. 
To put this into context, BaTiO$_{3}$ reaches a maximum shift current of 
6$\cdot$10$^{-3}$ nA~\cite{koch1975bulk,young_first_2012}.
As for the measurements performed in monolayer
WS$_{2}$, Ref.~\onlinecite{zhang_enhanced_2019}
did not report a photocurrent larger than 0.1~nA. 
This suggests that either our calculations overestimate the
peak shift-current value by at least an order of magnitude,
or some other effect counteracts its contribution.
Such effect could be the so-called
\emph{ballistic current}~\cite{belinicher1982kinetic,fridkin_bulk_2001}, and extrinsic kinetic 
contribution to the BPVE that can be as large or even 
larger than the shift current~\cite{sturman_ballistic_2020}.
%Many-body effects can also play an important role
%in determining the magnitude of quadratic light-absorption
%processes~\cite{PhysRevB.82.235201,doi:10.1073/pnas.1906938118,PhysRevB.107.205101}.
In any case, except for the peak magnitude our results on the monolayer 
are not inconsistent with the experimental 
findings of Ref.~\onlinecite{zhang_enhanced_2019},
given that in most spectral regions our calculations predict a shift current
smaller than 0.1 nA.

%Therefore, our analysis shows that while a WS$_{2}$ monolayer
%has a large shift photoconductivity of order $\simeq10$~$\mu$A/V$^{2}$, 
%the generated d.c. photocurrent is not particularly large
%under common experimental conditions. 
%The two main reasons are the relatively 
%small conducting area and the
%low light-absorption capability of the monolayer.

We come next to the nanotube results shown in Fig.~\ref{fig7}(b);
in our calculations, we have disregarded interwall interactions and 
employed the photoconductivity of 
a single-wall nanotube with $r=60$~\AA, together with $d=300$~\AA~in 
Eq.~\ref{eq:totalI} corresponding to a typical 
nanotube composed of $N=25$ layers~\cite{zak2010scaling,zhang_enhanced_2019}.
The calculated nanotube photocurrent in Fig.~\ref{fig7}(b) 
ranges between
order 1~nA at 2~eV and 10~nA at 2.9 eV, showing an elongated shape
around $\theta=0$ owing to the dominance of $\sigma^{xxx}$ 
over $\sigma^{xyy}$ (see Fig.~\ref{fig2}).
The maximum photocurrent 
measured in Ref.~\onlinecite{zhang_enhanced_2019}
is also of the order of 10~nA, but takes place at 2~eV, coinciding with the energy of 
the so-called A exciton of WS$_{2}$.
Aside from the mismatch in energy, our calculations show
that a zigzag WS$_{2}$ nanotube 
with shift photoconductivity 
equal to the monolayer
can  account for the order of magnitude measured in 
Ref.~\onlinecite{zhang_enhanced_2019}.
As for the angular dependence, the measured data for the 
nanotube shows 
a rounded shape around the origin,
but this distribution appears to depend significantly
on the precise nanotube that is measured~\cite{zhang_enhanced_2019}.

We note that  
the nanotubes used in experiment are typically 
composed of a mixture of internal structures
(\textit{i.e.} zigzag, armchair and chiral) that is in general unknown,
and even the radius is not constant throughout the whole 
nanotube, hence significant deviations from our idealized results are to be expected.
Improved theoretical results could be obtained by the modelling of 
interfaces between different types of structures.
Considering many-body interactions in the quadratic photoresponse~\cite{PhysRevB.82.235201,PhysRevB.103.075402,doi:10.1073/pnas.1906938118,huang_large_2023,PhysRevB.107.205101} 
could also bring numerical results closer to experiment. 
In particular, 
TMDs host the so-called A and B excitons~\cite{mccreary_-_2018}
that translate into narrow peaks in the spectra
(see Fig.~\ref{figRA}) which are not captured in our current
theoretical description.
The few available theoretical works reporting excitonic contributions
to the shift current indicate that the effect 
can be significant~\cite{PhysRevB.97.205432,PhysRevB.103.075402,doi:10.1073/pnas.1906938118,huang_large_2023}.

We expect to tackle these aspects in future work.
%on shift current which involves various methods, such as tight-binding~\cite{PhysRevB.103.075402}, GW+ Bethe Salpeter equation~\cite{huang_large_2023}, DFT+ non-equilibrium Green's function~\cite{doi:10.1073/pnas.1906938118} etc. %The current scope of this work is not taking such effects into account.

%In particular, different internal structures
%could tend to boost the photocurrent due to the internal electric field
%generated at the interfaces between two different structures.
%\pggt{We do not have excitons theory level, mention}

%As a final remark, we note that the photocurrent
%measured in Ref.~\onlinecite{zhang_enhanced_2019} varied by
%as much as two orders of magnitude between
%different nanotube devices. 
%A real nanotube is composed of a mixture of internal structures
%(i.e. zigzag, armchair and chiral) that is in general unknown,
%and even the radius is not constant throughout the whole 
%nanotube. 
%These deviations from the idealized structure
%Aside from contributions such as the ballistic current that we did 
%not consider in the present work, 

\section{Conclusions}
\label{sec:conclusions}

In summary, we have 
conducted a systematic study of the shift current
in WS$_{2}$ monolayer and nanotube structures.
Our DFT calculations have shown that 
the optical properties of a single wall zigzag nanotube 
are well described by those of
a monolayer for nanotube radius larger than $\sim60$~\AA. 
According to our calculations,  the single-wall results are only
slightly modified when accounting for  
interactions with other walls of the nanotube for typical interwall distances.
Despite possessing a similar shift photoconductivity,
we have shown that a WS$_{2}$ 
nanotube can generate a photocurrent of 
around 10~nA, while the monolayer attains a maximum
photocurrent of order 1~nA.
The main reason behind this difference is the  
larger conducting cross section of a nanotube in comparison to a monolayer.
Our calculations reproduce the 
order of magnitude of the photocurrent
measured in a recent experiment
on WS$_{2}$ nanotubes~\cite{zhang_enhanced_2019}, 
suggesting that the shift current plays an important role.

%Finally, we have rationalized our results in the context 
%of recent experimental measurements~\cite{zhang_enhanced_2019}. 

%Ballistic current might hinder current generation, specially in
%monolayer.
%Inner structure of nanotube is largely unknown (mix of 
%zigzag, armchair, chiral ...), full of interfaces, even number
%of radius is not fixed. Shift current is closely tied to symmetry
%breaking, this might enhance shift current generation. 

\section{Acknowledgments}
We are very grateful to Yijin Zhang for stimulating correspondence. 
This project has received funding from the European Union’s Horizon 2020 research and innovation programme under the European Research Council (ERC) Grant Agreement No. 946629, 
and the Department of Education, Universities and Research 
of the Eusko Jaurlaritza and the University of the 
Basque Country UPV/EHU (Grant No. IT1527-22).
%and the Department of Education, Universities and Research 
%of the Eusko Jaurlaritza and the University of the 
%Basque Country UPV/EHU (Grant No. IT1527-22).

\appendix
\section*{Appendix: additional calculations}
\label{sec:appendix}
Here we provide additional calculations on electronic structure and 
optical properties. 

\renewcommand{\thefigure}{A\arabic{figure}}
\setcounter{figure}{0}  

\subsection{Band structure and Wannier interpolation} 

\begin{figure}[h]
    \centering
    \includegraphics[width=0.5\textwidth]{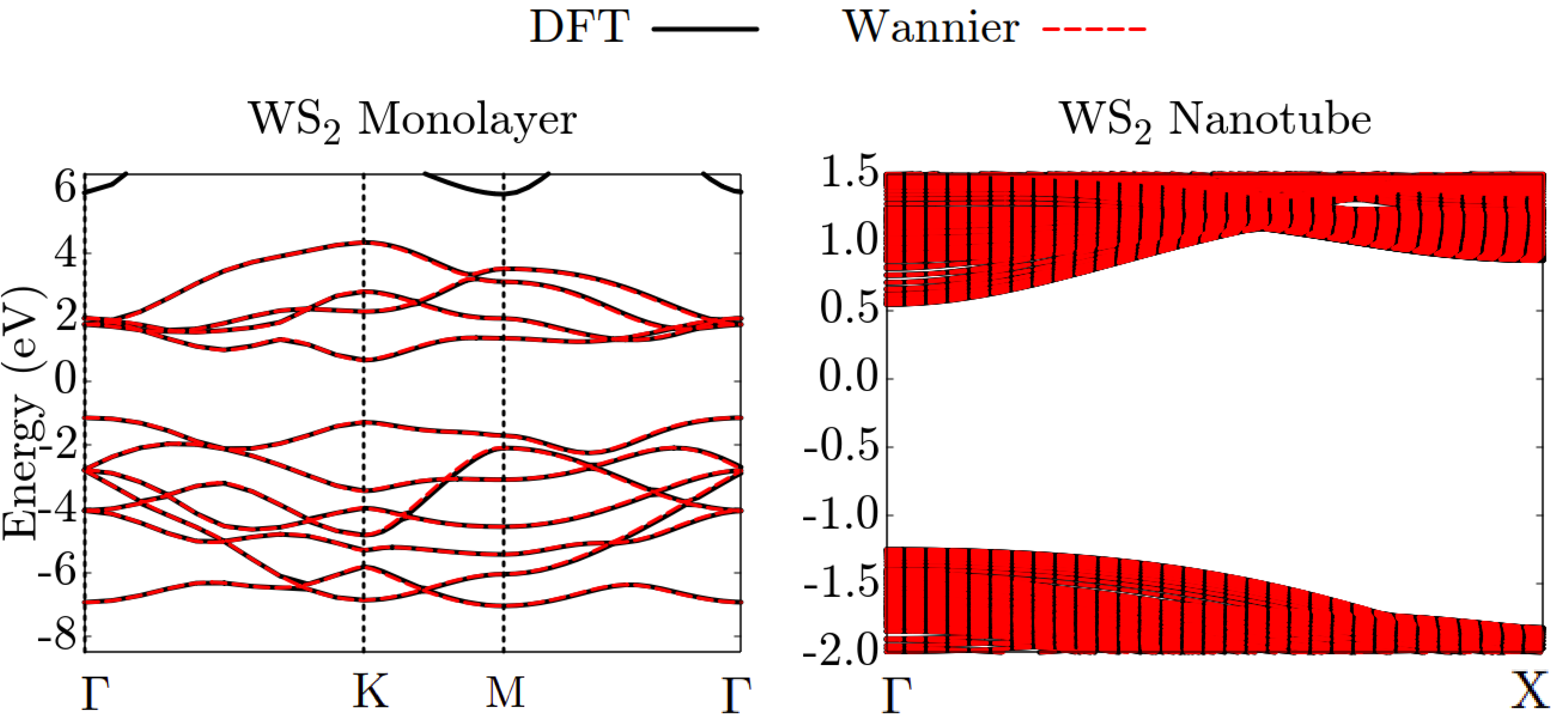}
      \caption{The DFT (in solid black line) and Wannier-interpolated (in dashed red line) bandstructure of monolayer and nanotube. The Fermi energy level is set at 0 eV.}
    \label{figA1}
\end{figure}

Fig.~\ref{figA1} shows the calculated band structure for WS$_{2}$
monolayer [Fig.~\ref{fig1}(b)] and zigzag nanotube of radius $r=40$~\AA~ [Fig.~\ref{fig1}(a)].
We have also included the Wannier-interpolated band structure, which
reproduces the DFT one. 
The figure shows that the direct band gap of the monolayer takes
place at high symmetry point K, while in the nanotube it takes place
at $\Gamma$. The value of the band gap is $\simeq2$ eV, virtually the same
in both structures

%%%%%%%%FIGURE A1 %%%%%%%%%%%%%%%%%%%%
%%%%%%%%%%%%%%%%%%%%%%%%%%%%%%%%%%%%%

\subsection{Shift current} 
\subsubsection{Atomic relaxation}
In the calculations in the main text
we considered ideal (unrelaxed) 
atomic positions for the nanotubes.
The reason to proceed in this way is twofold. 
Firstly, we have checked that the relaxation procedure 
results in changes of the bond lengths
below 1~\%;  this implies that the shift current is only mildly affected as compared to the ideal structure, as exemplified in Fig.~\ref{figA2-4}(a) for a zigzag nanotube of $r=40$~\AA. 
In all calculations, the computed static stress is $<$ 0.01 eV$/$~\AA$^{3}$.
%In this way we avoid relaxing the largest nanotubes, which
%is computationally expensive. 
Secondly, working with the ideal atomic structure makes sure that
the monolayer and nanotube geometries are as close as possible.
This then allows a clear comparison of the optical properties of a monolayer and the corresponding nanotube constructed from it, 
better highlighting the similarities and differences between them.

\subsubsection{Spin-orbit coupling}
Spin-orbit coupling (SOC) is usually not the main driving effect 
for the shift current of semiconductors. However, given that 
tungsten is a heavy element, we
have conducted additional calculations on the shift photoconductivity
including SOC. We have conducted these calculations 
using the plane-wave-based \textsc{QUANTUM ESPRESSO} code package~\cite{giannozzi2017advanced}, given that
the interface to the \textsc{Wannier90} code package is currently implemented
for the case of fully relativistic pseudopotentials. 
Fig.~\ref{figA2-4}(b) shows the dominant tensor component 
of the shift photoconductivity $\sigma_{2}^{xxx}$
of monolayer WS$_{2}$ calculated with and without SOC.
The figure shows that the band-gap energy is reduced by 
$\simeq0.2$ eV as a consequence of SOC, and the main spectral 
features are somewhat shifted to lower energies roughly 
by that amount.
As expected, the overall order of magnitude of the shift photoconductivity 
is not altered by SOC. We have verified that this is also the case for 
calculations  of nanotube structures, 
and that the main results of this work are
also not modified by SOC.

\subsubsection{Nanotube configurations}
In addition to the zigzag configuration, an experimentally synthesized TMD nanotube can coexist with two other configurations, namely armchair $(n,n)$ and chiral $(n,m)$. In Fig.~\ref{figA2-4}(c), we present a comparative analysis of the dominant tensor of the shift photoconductivity $\sigma_{2}^{xxx}$ of three different WS$_{2}$ nanotube configurations, namely zigzag, armchair and chiral, each with a radius of 30~{\AA}. The figure shows that  $\sigma_{2}^{xxx}$ for the armchair configuration 
is two orders of magnitude smaller as compared to the zigzag. On the other hand, 
the shift current of 
chiral nanotubes follows the same peak 
trend and same order of magnitude as the zigzag.

\subsubsection{Different TMDs}
\begin{figure*}[t]
    \centering
    \includegraphics[width=1.0\textwidth]{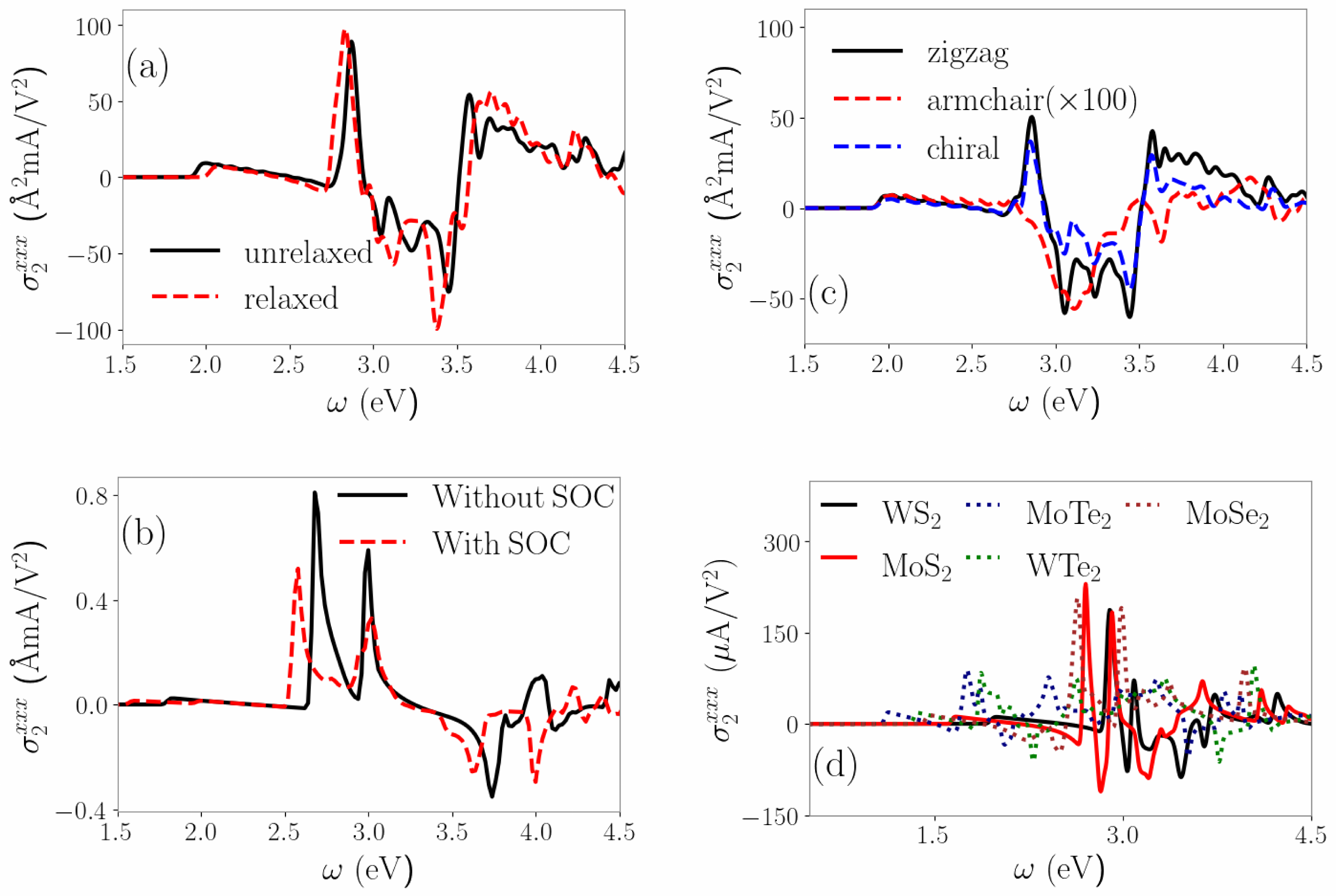}
      \caption{
      Dominant tensor component of shift photoconductivity $\sigma^{xxx}_{2}$
      as a function of frequency for various cases.      
      (a) Unrelaxed (black line) and relaxed (red line) structure of a zigzag WS$_{2}$
      nanotube of $r=40$~{\AA}. (b) Calculations in monolayer WS$_{2}$ 
      with and without spin-orbit coupling as shown in red dashed and black solid line, respectively. (c) Zigzag (in black solid), armchair (in red dashed), and chiral (in blue dashed) nanotube configurations of $r=30$~{\AA}. The chiral angle for the chiral nanotube is 16.102$^{\circ}$. (d) Several MX$_{2}$ combinations (M = Mo/W; X = S, Te, Se).}
    \label{figA2-4}
\end{figure*}

Finally, we have extended the calculations to cover the TMD combinations  $M=$Mo, W and $X=$S, Se, Te. 
Fig.~\ref{figA2-4}(d) shows the calculated $\sigma_{2}^{xxx}$  (including SOC)
of the monolayer structure for all these combinations. 
As shown by the figure, the overall order of magnitude
is the same for all the TMDs, with a maximum 
shift photoconductivity of 220 $\upmu$A/V$^{2}$ attained in 
MoS$_{2}$. 

 %%%%%%%%FIGURE A2-4 %%%%%%%%%%%%%%%%%%%%
%%%%%%%%%%%%%%%%%%%%%%%%%%%%%%%%%%%%%
\newpage

%\bibliography{paper}

\end{document}